\documentclass[aps,prl,twocolumn,preprintnumbers,superscriptaddress,10pt]{revtex4-1}
\usepackage[mathscr]{eucal}
\usepackage{color,amsmath,amssymb,amsfonts}
\usepackage{enumerate}
\usepackage{hyperref}
\usepackage{graphicx}
\usepackage{psfrag}
\usepackage{dcolumn,bm}
\usepackage{mdframed}

\usepackage[utf8]{inputenc}

\newcommand{\bea}{\begin{eqnarray}}
\newcommand{\beal}[1]{\begin{eqnarray}\label{#1}}
\newcommand{\eea}{\end{eqnarray}}
\newcommand{\be}{\begin{equation}}
\newcommand{\bel}[1]{\begin{equation}\label{#1}}
\newcommand{\ee}{\end{equation}}

\newcommand{\bit}{\begin{itemize}}
\newcommand{\eit}{\end{itemize}}
\newcommand{\ben}{\begin{enumerate}}
\newcommand{\een}{\end{enumerate}}

\def\del{\partial}

\newcommand{\mt}[1]{\textrm{\tiny #1}}
\newcommand{\nc}{N_\mt{c}}

\newcommand{\sac}{\, , \qquad}

\newcommand{\thy}{t_\textrm{hyd}}
\newcommand{\teq}{t_\textrm{eq}}
\newcommand{\Thy}{T_\textrm{hyd}}
\newcommand{\Teq}{T_\textrm{eq}}

\newcommand{\eqq}[1]{(\ref{#1})}
\newcommand{\fig}[1]{Fig.~\ref{#1}}

\newcommand{\bal}{\begin{align}}
\newcommand{\eal}{\end{align}}
\newcommand{\bse}{\begin{subequations}}
\newcommand{\ese}{\end{subequations}}

\def\P{{{P}}}

\def\Peq{{{P}}_{\rm eq}}

\begin{document}
\onecolumngrid

\preprint{ICCUB-18-013}

\title{Holographic Collisions across a Phase Transition}

\author{Maximilian Attems}
\affiliation{Instituto Galego de F\'\i sica de Altas Enerx\'\i as (IGFAE), Universidade de Santiago de Compostela, 15782 Galicia, Spain}
\affiliation{Departament de F\'\i sica Qu\`antica i Astrof\'\i sica \&  Institut de Ci\`encies del Cosmos (ICC), Universitat de Barcelona, Mart\'{\i}  i Franqu\`es 1, 08028 Barcelona, Spain}
\author{Yago Bea}
\affiliation{Departament de F\'\i sica Qu\`antica i Astrof\'\i sica \&  Institut de Ci\`encies del Cosmos (ICC), Universitat de Barcelona, Mart\'{\i}  i Franqu\`es 1, 08028 Barcelona, Spain}
\author{Jorge Casalderrey-Solana}
\affiliation{Departament de F\'\i sica Qu\`antica i Astrof\'\i sica \&  Institut de Ci\`encies del Cosmos (ICC), Universitat de Barcelona, Mart\'{\i}  i Franqu\`es 1, 08028 Barcelona, Spain} 
\affiliation{Rudolf Peierls Centre for Theoretical Physics, University of Oxford, Clarendon Laboratory, Parks Rd, Oxford OX1 3PU, United Kingdom,}
\author{David Mateos}
\affiliation{Departament de F\'\i sica Qu\`antica i Astrof\'\i sica \&  Institut de Ci\`encies del Cosmos (ICC), Universitat de Barcelona, Mart\'{\i}  i Franqu\`es 1, 08028 Barcelona, Spain}
\affiliation{Instituci\'o Catalana de Recerca i Estudis Avan\c cats (ICREA), 
Llu\'\i s Companys 23, Barcelona, Spain}
\author{Miquel Triana}
\affiliation{Departament de F\'\i sica Qu\`antica i Astrof\'\i sica \&  Institut de Ci\`encies del Cosmos (ICC), Universitat de Barcelona, Mart\'{\i}  i Franqu\`es 1, 08028 Barcelona, Spain}
\author{Miguel Zilh\~ao}
\affiliation{CENTRA, Departamento de F\'\i sica, Instituto Superior T\'ecnico, Universidade de Lisboa, Avenida Rovisco Pais 1, 1049 Lisboa, Portugal}
\affiliation{Departament de F\'\i sica Qu\`antica i Astrof\'\i sica \&  Institut de Ci\`encies del Cosmos (ICC), Universitat de Barcelona, Mart\'{\i}  i Franqu\`es 1, 08028 Barcelona, Spain}


\begin{abstract}
We use holography to analyse relativistic collisions in a one-parameter family of strongly coupled gauge theories with thermal phase transitions. For a critical value of the parameter the transition is second order, for subcritical values it is first order, and for super-critical values it a smooth crossover. We extract the gauge theory stress tensor from collisions of gravitational shock waves on the dual geometries. Regardless of the nature of the transition, for  values of the parameter close to the critical value almost all the energy of the projectiles is deposited into a long-lived, quasi-static blob of energy at mid-rapidity. This configuration is well described by the constitutive relations of second-order hydrodynamics that include all second-order gradients that are purely spatial in the local rest frame. In contrast, a M\"uller-Israel-Stewart-type formulation of hydrodynamics 
fails to provide a good description. 
We discuss possible implications for searches of the QCD critical point.
\end{abstract}
\maketitle

\noindent
{{\bf 1. Introduction.}}
``Holographic Collisions'', namely collisions of gravitational shockwaves in an asymptotically AdS spacetime, have provided interesting insights into the far-from-equilibrium properties of hot, strongly-coupled, non-Abelian plasmas that are potentially relevant for the quark-gluon plasma (QGP) created in heavy ion collision experiments (see e.g.~\cite{CasalderreySolana:2011us} for a review). The first examples \cite{Chesler:2010bi,Casalderrey-Solana:2013aba,Casalderrey-Solana:2013sxa,Chesler:2015wra,Chesler:2015bba,Chesler:2015lsa,Chesler:2016ceu} considered gravity models dual to conformal field theories (CFTs). These studies were subsequently extended to non-conformal theories in \cite{Attems:2016tby,Attems:2017zam} based on the set of models introduced in \cite{Attems:2016ugt}. 

The purpose of this paper is to perform the first simulations of holographic collisions in theories with thermal phase transitions; previous holographic studies of time evolution in theories with phase transitions include \cite{Gursoy:2016ggq,Attems:2017ezz,Janik:2017ykj,Critelli:2018osu}. Our main motivation is that, if  QCD possesses a critical point, future collision experiments, such as the BES-II program at RHIC, the CBM experiment at FAIR or upcoming experiments at NICA,  may probe the out-of-equilibrium dynamics of the QGP across a phase transition. We will show that the formulation of hydrodynamics that could potentially describe this dynamics is not the formulation that is widely  used in hydrodynamic codes \cite{Romatschke:2017ejr}. While the consequences of this conclusion are far-reaching, fortunately our results also suggest the direction in which these codes may need to be modified in order to capture the correct physics.  

\noindent
{{\bf 2. The model.}} 
Our gravity model is described by the 
Einstein-scalar action
\begin{equation}
\label{eq:action}
S=\frac{2}{\kappa_5^2} \int d^5 x \sqrt{-g} \left[ \frac{1}{4} {\cal R}  - \frac{1}{2} \left( \nabla \phi \right) ^2 - V(\phi) \right ] ~,
\end{equation}
with potential 
\begin{equation}
\ell^2 V(\phi)=-3-\frac{3\phi^2}{2}-\frac{\phi^4}{3}-\frac{\phi^6}{3\phi_M^2}+\frac{\phi^6}{2\phi_M^4}-\frac{\phi^8}{12\phi_M^4}~,
\label{potential}
\end{equation}
with $\ell$ a length scale and $\phi_M$ a free parameter.  This is exactly the potential of \cite{Attems:2017ezz}. It has a maximum at $\phi=0$, where it admits an exact AdS solution of radius $\ell$. The dual gauge theory is a CFT deformed by a source $\Lambda$ for the dimension-three scalar operator $\mathcal{O}$ dual to the scalar field $\phi$. Additional details may be found in 
\cite{Attems:2017ezz,inprogress}.

\begin{figure*}[t!!!]
\begin{center}
\begin{tabular}{cc}
\includegraphics[width=.3\textwidth]{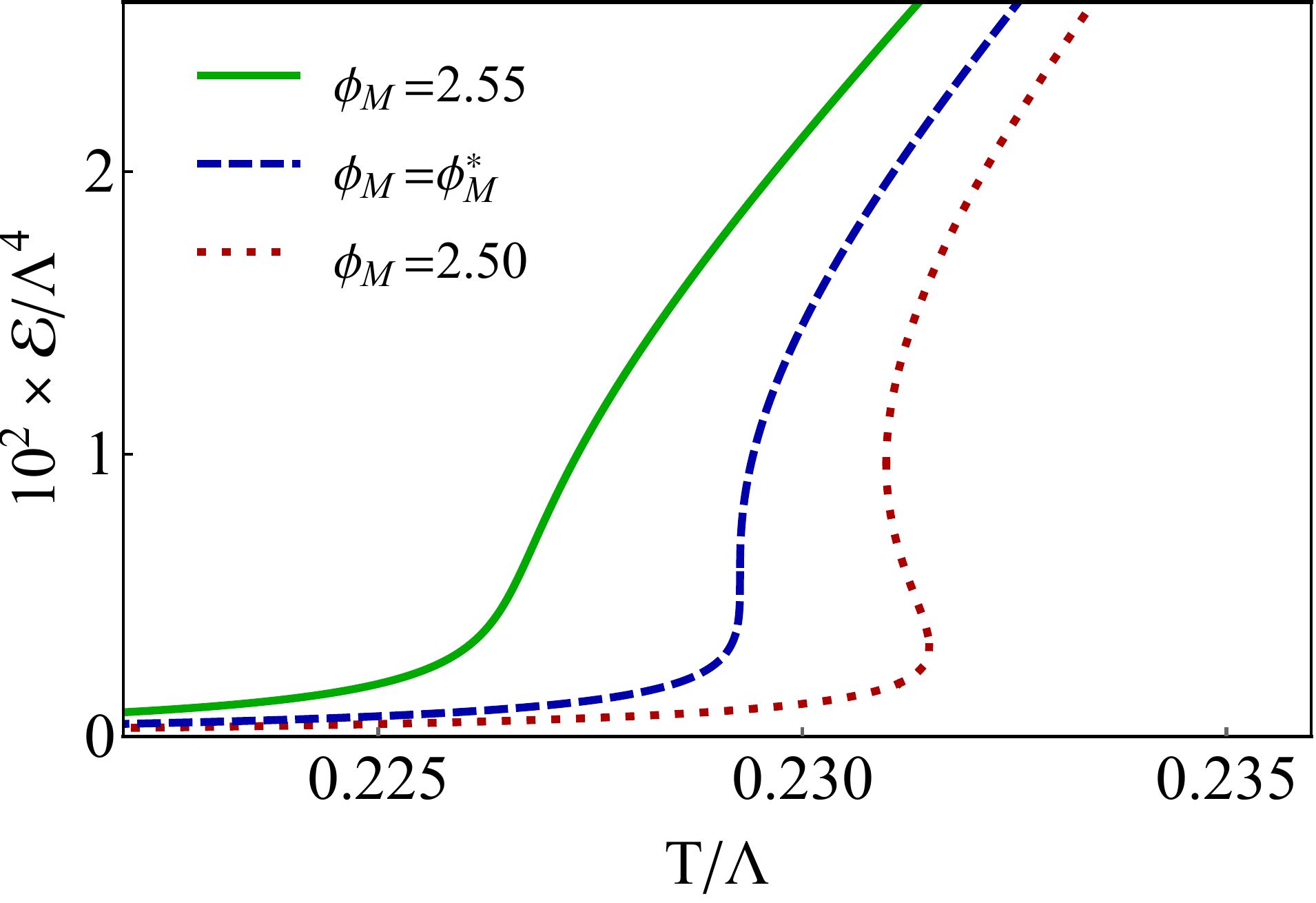} 
\,&\,
\includegraphics[width=.3\textwidth]{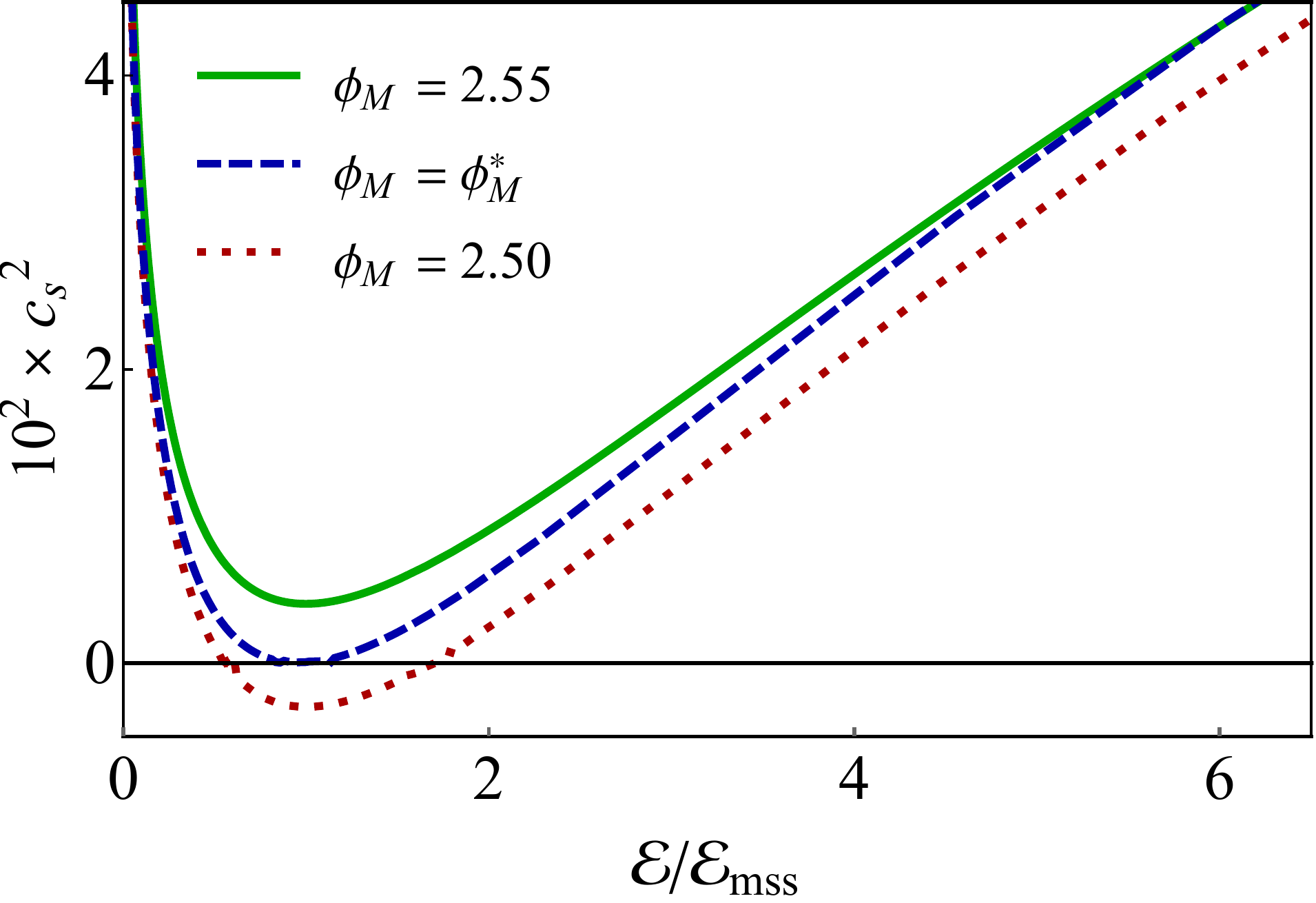} 
\end{tabular}
\end{center}
\vspace{-5mm}
\caption{\label{fig:energyVStemperature0} \small
Energy density  (left) and speed of sound squared  (right) for different theories. $\mathcal{E}_\textrm{mss}$ is the energy density with the minimal value of $c_s^2$, which is given by 
$10^3 \mathcal{E}_\textrm{mss} = \{5.7, 5.8, 5.4 \}$ for $\phi_M=\{2.50, \phi_M^*, 2.55 \}$, respectively.
}
\vspace{-3mm}
\end{figure*}

The thermodynamics of the theory depends crucially on the value of $\phi_M$, as shown by the plots in \fig{fig:energyVStemperature0}. For the critical value $\phi_M^*\simeq 2.521$ the theory possesses a second-order phase transition and the speed of sound squared, $c_s^2$, vanishes at the critical temperature.  For sub-critical values $\phi \lesssim \phi_M$ the transition is first-order and $c_s^2$ attains negative values. For super-critical values $\phi \gtrsim \phi_M$ the transition is actually a smooth crossover and $c_s^2$ attains small but always positive values. 

\noindent
{{\bf 3. Collisions.}} We collide gravitational shock-waves in the model \eqq{eq:action} as described in~\cite{Attems:2016tby,Attems:2017ezz}. We use  ``1/2-shocks'' in the language of \cite{Casalderrey-Solana:2013aba}. We choose the energy of the projectiles so that the post-collision evolution explores the transition region of the gauge theory phase diagram.  
\begin{figure*}[!htb]
\vspace{5mm}
\begin{center}
\begin{tabular}{ccc}
\includegraphics[width=.31\textwidth]{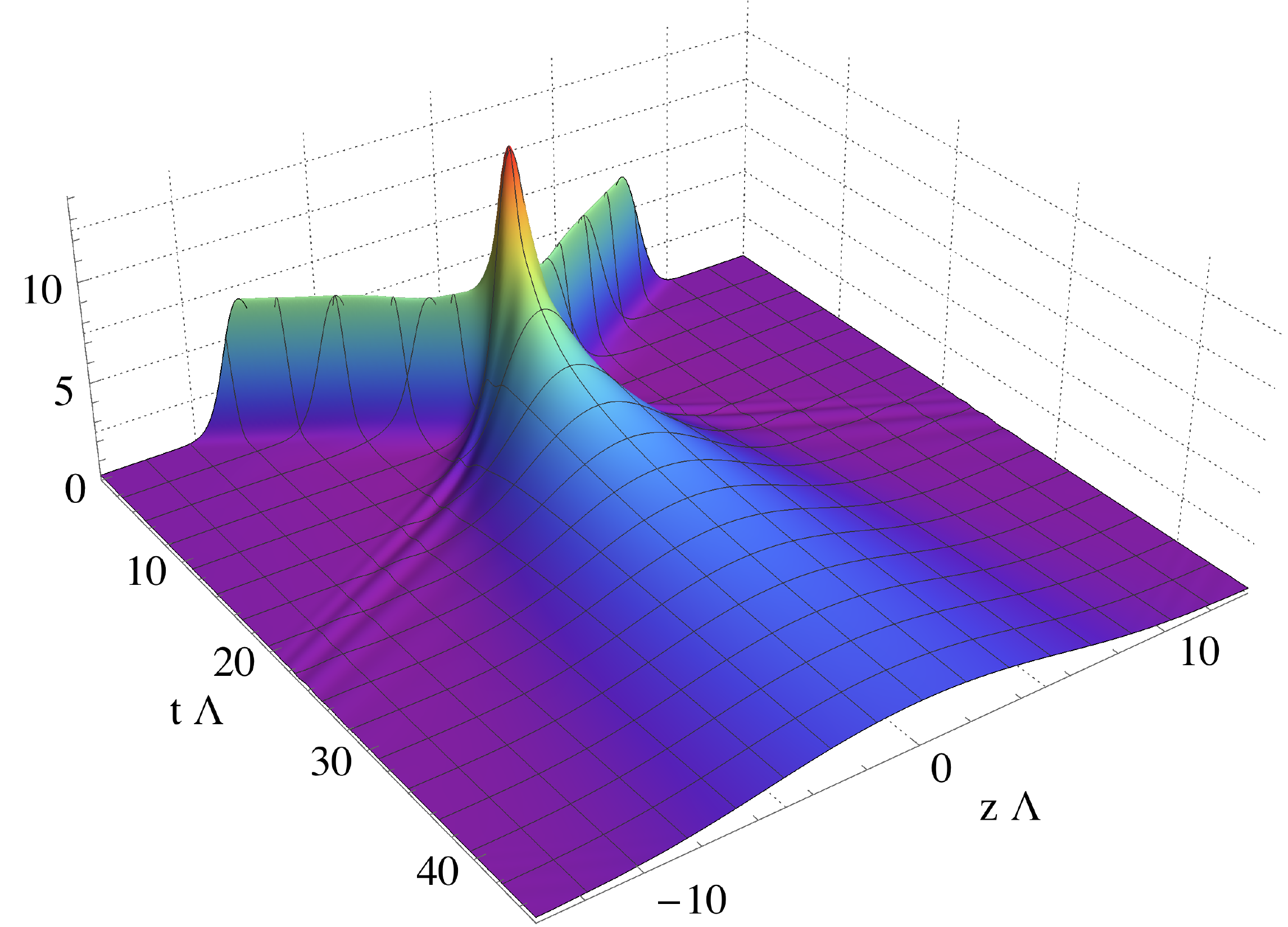} 
\put(-165,95){\mbox{$\mathcal{E}/\mathcal{E}_\textrm{mss}$}}
\put(-90,125){\mbox{$\phi_M=2.50$}}
\quad&\quad
\includegraphics[width=.31\textwidth]{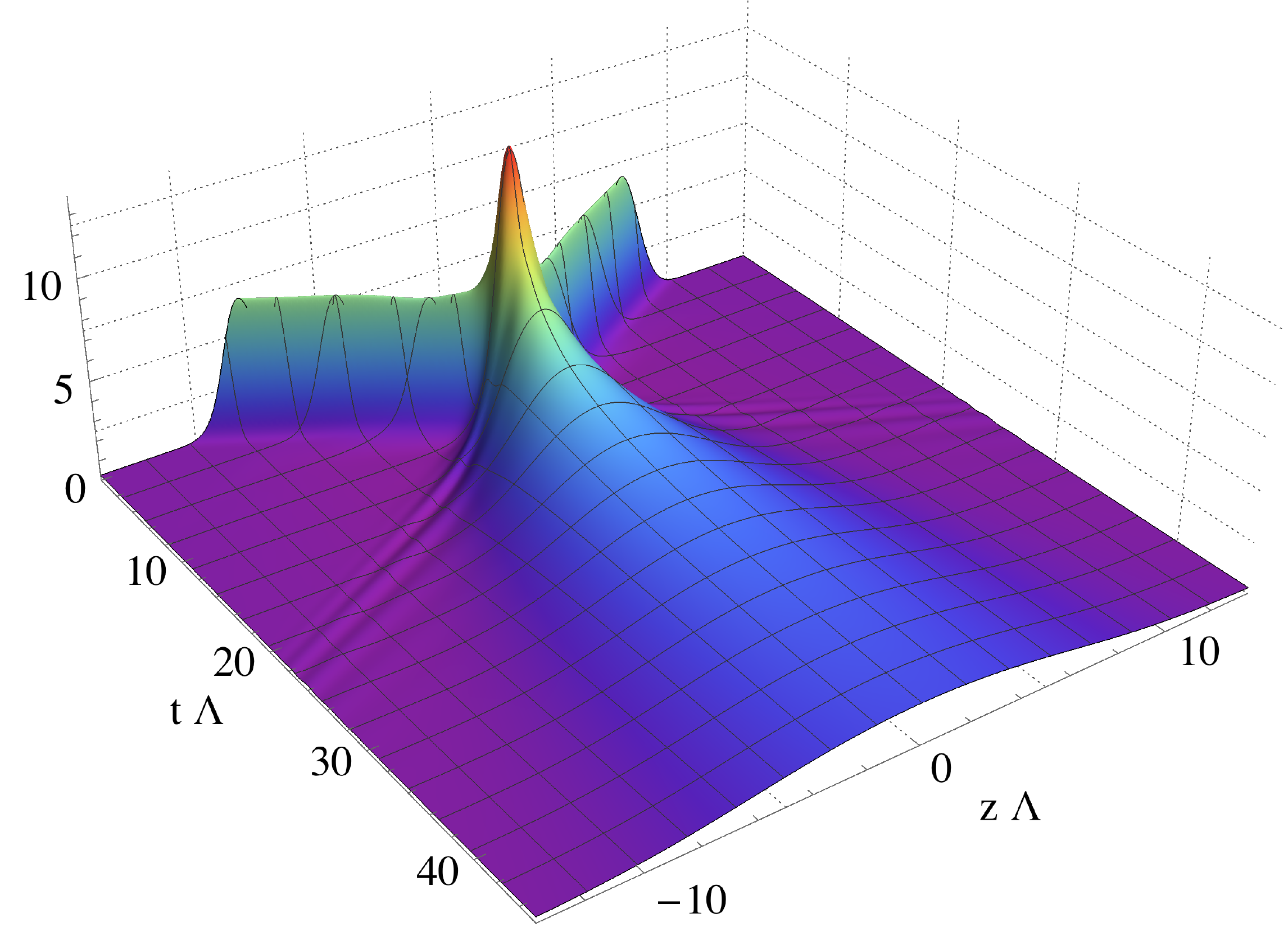} 
\put(-165,95){\mbox{$\mathcal{E}/\mathcal{E}_\textrm{mss}$}}
\put(-90,125){\mbox{$\phi_M=\phi_M^*$}}
\quad&\quad
\includegraphics[width=.31\textwidth]{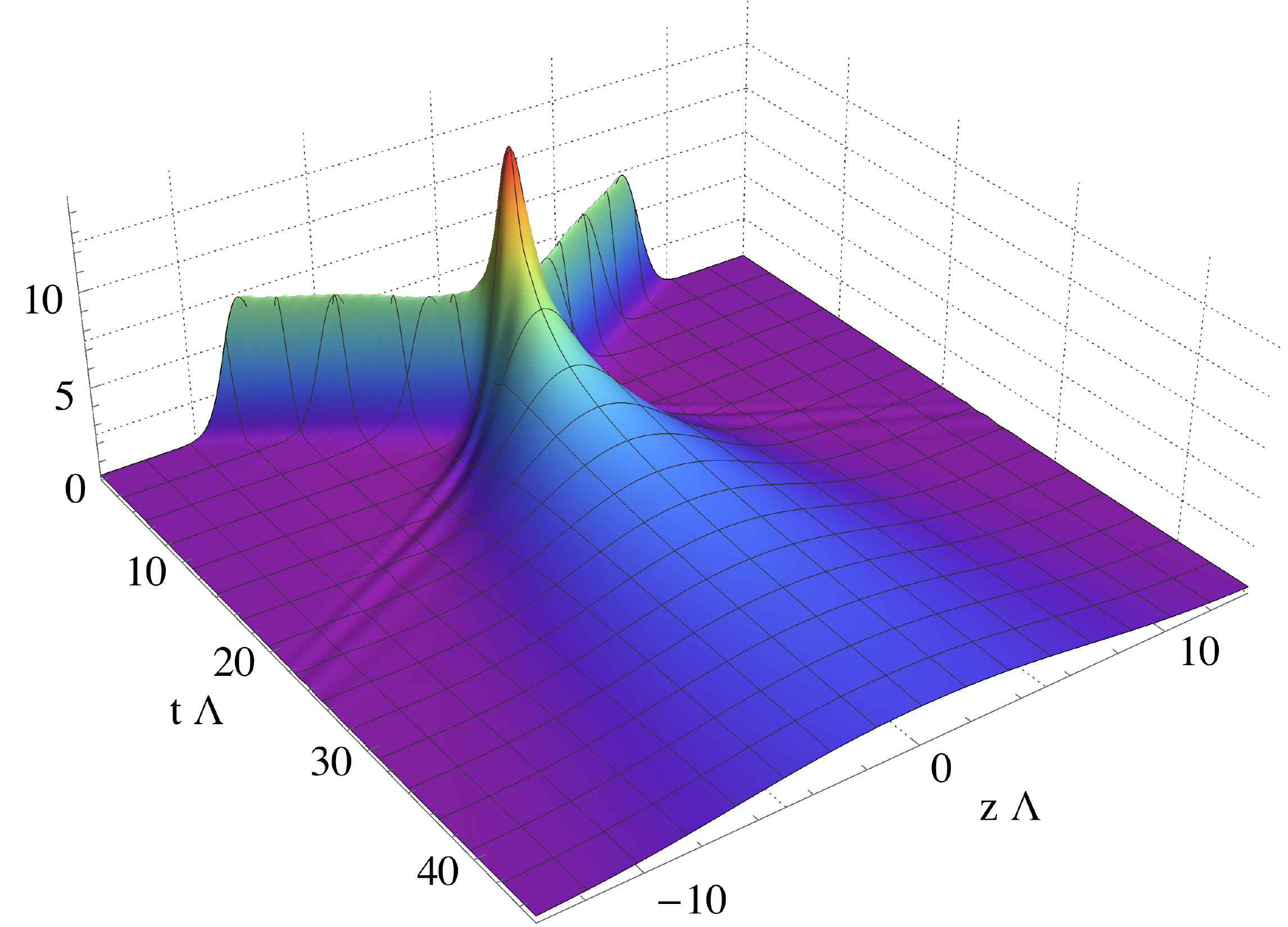}
\put(-165,95){\mbox{$\mathcal{E}/\mathcal{E}_\textrm{mss}$}}
\put(-90,125){\mbox{$\phi_M=2.55$}}
\end{tabular}
\end{center}
\caption{\label{fig:EnergyScanCriticalPoint} \small Spacetime evolution of the gauge theory energy density resulting from the collisions.
}
\end{figure*}
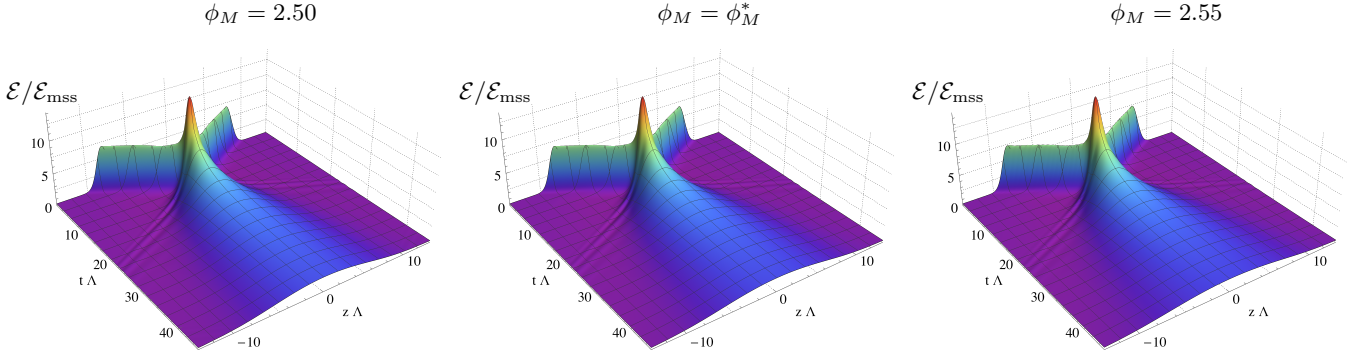 

The result of the collisions is shown in \fig{fig:EnergyScanCriticalPoint}, where we plot the gauge theory energy density as a function of time $t$ and of position along the collision direction $z$. The main two lessons are as follows. First, most of the energy of the incoming projectiles is deposited into a long-lived, quasi-static blob of energy at mid-rapidity. This in contrast to collisions in theories without phase transitions \cite{Chesler:2010bi,Casalderrey-Solana:2013aba,Casalderrey-Solana:2013sxa,Chesler:2015wra,Chesler:2015bba,Chesler:2015lsa,Chesler:2016ceu,Attems:2016tby,Attems:2017zam}, where the energy density profile after the collision exhibits a minimum at mid-rapidity and two maxima away from mid-rapidity. Second, the physics of the collision is qualitatively  identical for the three values of $\phi_M$ shown in \fig{fig:energyVStemperature0}, regardless of whether the equilibrium physics is that of a first-order transition, a second-order transition or a smooth crossover.

\noindent
{{\bf 4. Hydrodynamics.}} 
We will now show that the long-lived, quasi-static state formed in these collisions is well described by second-order hydrodynamics. We focus on the value $\phi_M=2.3$, for which the transition is first order, because in this case the relevant second-order transport coefficients can be extracted from the static, inhomogeneous configurations of \cite{Attems:2017ezz}. 

\begin{figure*}[!htb]
\begin{center}
\begin{tabular}{cc}
\includegraphics[width=.45\textwidth]{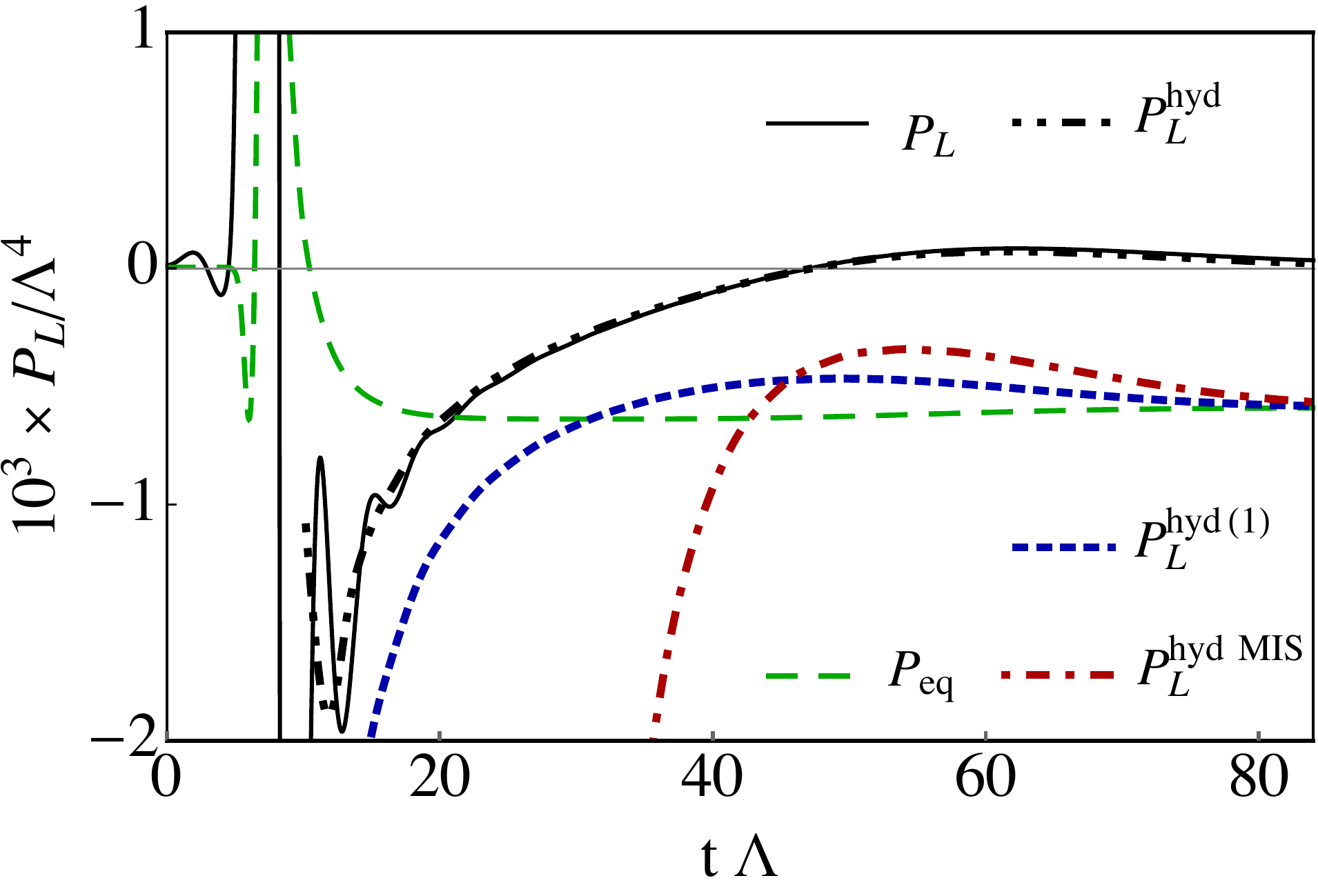} 
\,&\,
\includegraphics[width=.45\textwidth]{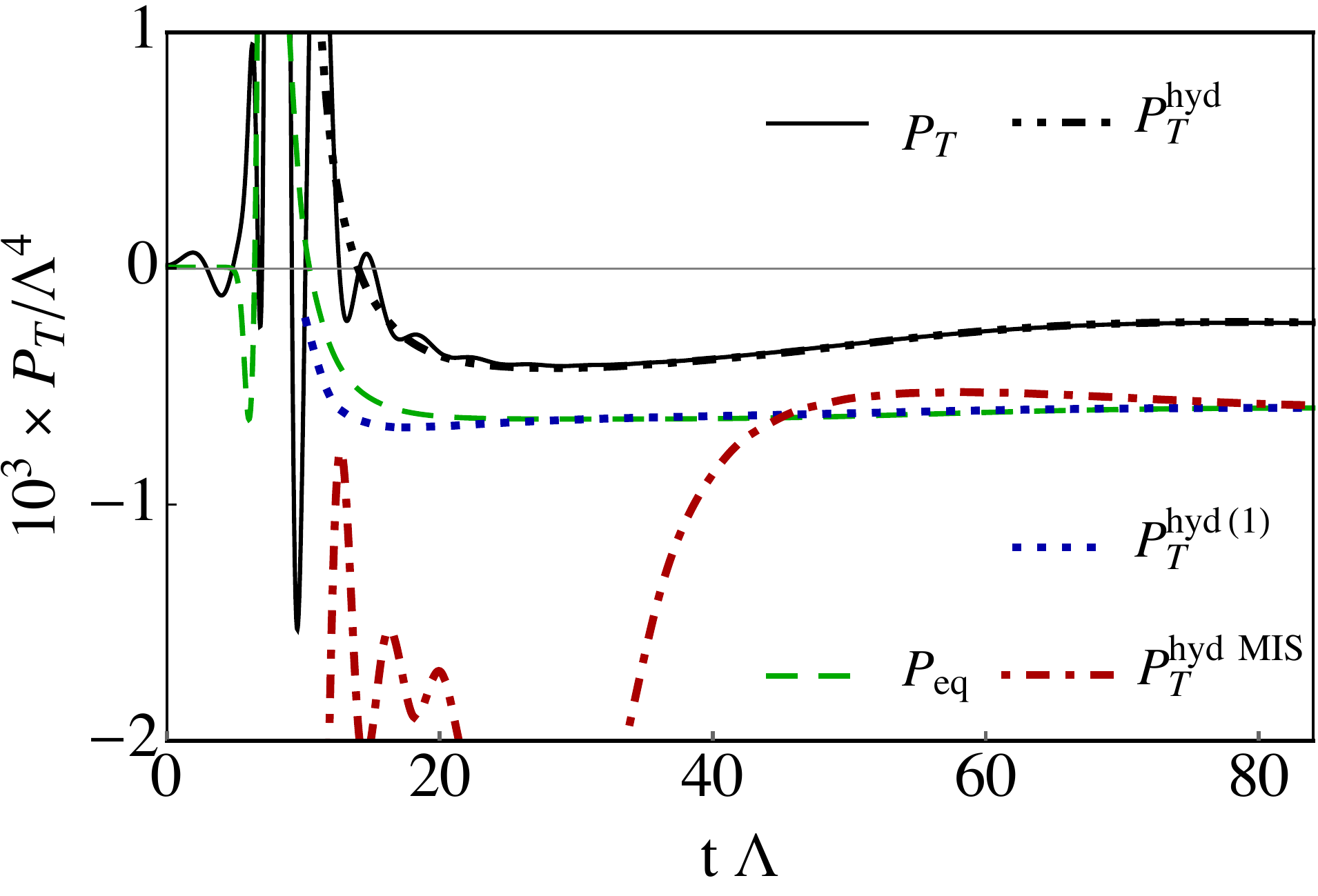}
\\
[4mm]
\includegraphics[width=.45\textwidth]{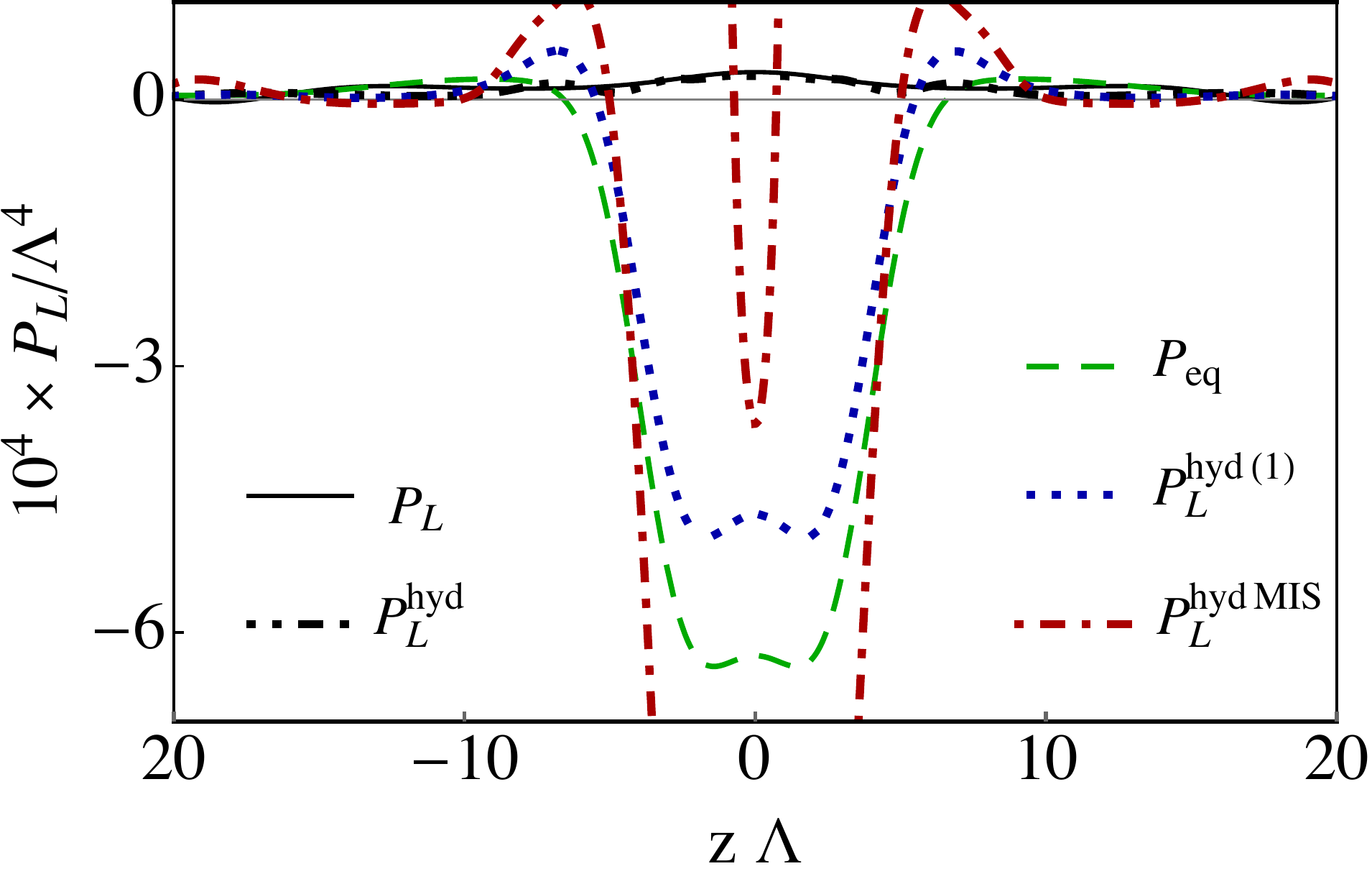} 
\,&\,
\includegraphics[width=.45\textwidth]{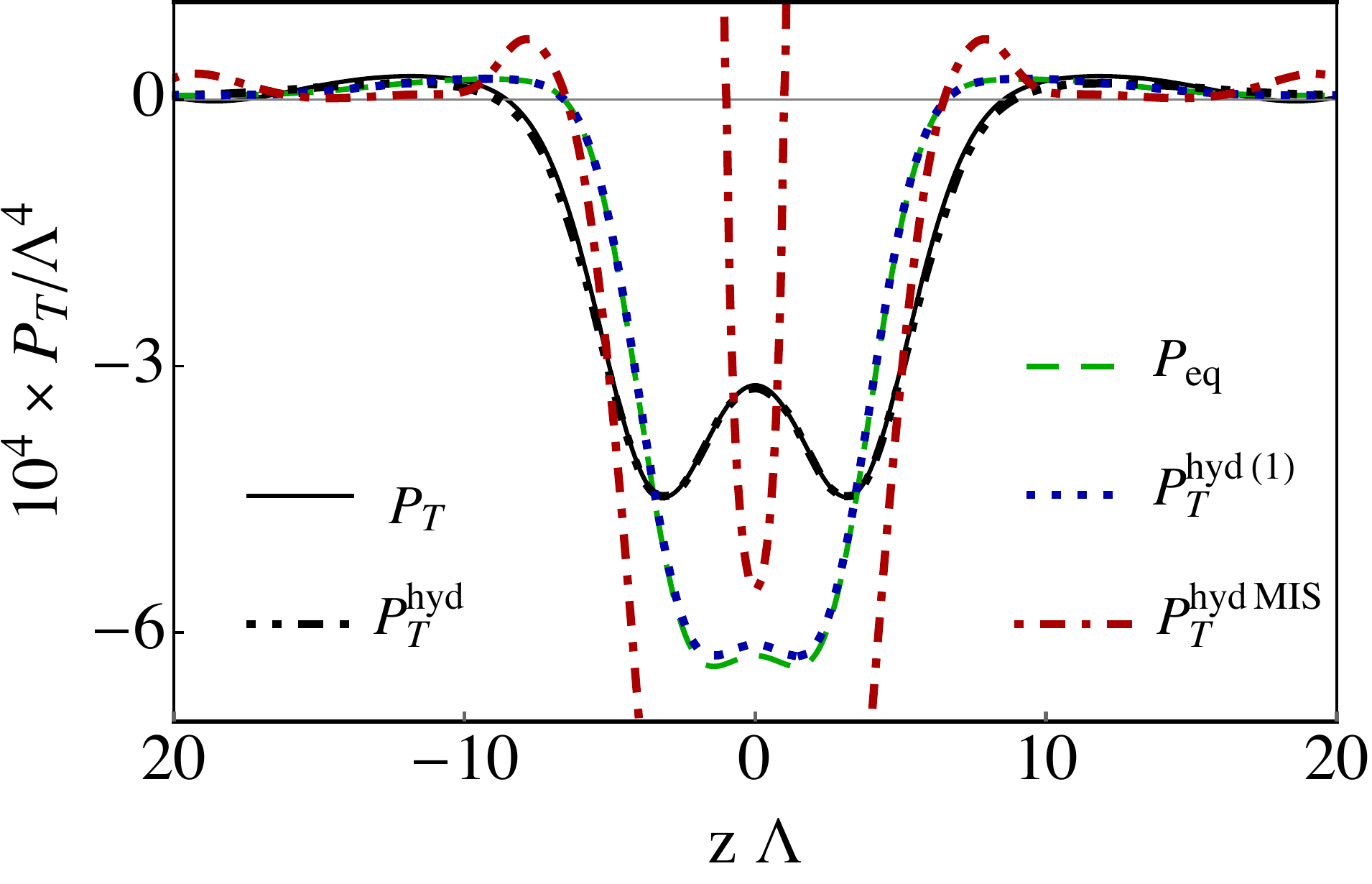} 
\end{tabular}
\end{center}
\vspace{-5mm}
\caption{\label{fig:pressures_midrapidity} \small
(Top row) Pressures at mid rapidity. (Bottom row) Snapshots of the pressures at $ t = 50/\Lambda > \thy$. $P_L, P_T$ are the exact pressures  extracted from gravity. $P_{\mt{eq}}$ is the equilibrium pressure. $P_L^{\mt{hyd(1)}}, P_T^{\mt{hyd(1)}}$ are the hydrodynamic pressures with only first-order viscous corrections included, i.e.~those in 
$T^{\mu \nu}_\mt{ideal} + \Pi_{(1)}^{\mu\nu} $.  $P_L^{\mt{hyd}}, P_T^{\mt{hyd}}$ are the  second-order hydrodynamic pressures in the purely spatial formulation omitting the contributions of 
$\tilde{\mathcal{O}}_7^{\mu\nu}$ and 
$\tilde{\mathcal{S}}_4$. 
$P_L^{\mt{hydMIS}}, P_T^{\mt{hydMIS}}$ are the  second-order hydrodynamic pressures in the MIS-type formulation.
}
\vspace{-3mm}
\end{figure*}

In modern language we define hydrodynamics as a gradient expansion around local equilibrium that, at any given order, includes all possible gradients of the hydrodynamic variables that are purely spatial in the local rest frame. Let us refer to this as the purely spatial formulation. 
To second order the hydrodynamic stress tensor takes the form 
\be
\label{stresstensor}
T^{\mu \nu} = T^{\mu \nu}_\mt{ideal} + \Pi^{\mu\nu}\sac 
\Pi^{\mu\nu} = \Pi_{(1)}^{\mu\nu} + \Pi_{(2)}^{\mu\nu} \,,
\ee
with
\bse
\label{omitted}
\bal
\label{bothterms}
\Pi_{(1)}^{\mu\nu} &= -\eta\sigma^{\mu\nu} -\zeta \left( \nabla\cdot u \right) \Delta^{\mu\nu} \,, \\[2mm]
\Pi_{(2)}^{\mu\nu} &= \pi_{(2)}^{\mu\nu} + \Delta^{\mu\nu} \Pi_{(2)} \,. 
\end{align}
\ese
In these expressions $u^\mu$ is the fluid four-velocity, $\Delta^{\mu\nu}$ is the projector onto spatial directions in the local rest frame, and 
$\Pi_{(1)}^{\mu\nu}$ contains the first-order corrections, with 
$\eta$ and $\zeta$  the shear and bulk viscosities, respectively. The shear tensor is  $\sigma^{\mu\nu} = \nabla^{<\mu} u^{\nu>} $, where $\nabla^\mu\equiv \Delta^{\mu\nu} \del_\nu$ and $A^{<\mu \nu >}$ denotes the symmetric, transverse and traceless part of any rank-two tensor. Note that all first-order terms are  linear in the velocity. As in other holographic models  as e.g.~\cite{DeWolfe:2011ts}, the bulk viscosity remains finite at the second-order phase transition as a consequence of the large-$N_c$ approximation implicit in the holographic set-up \cite{Natsuume:2010bs}.
 All the second-order terms are contained in $\Pi_{(2)}^{\mu\nu}$. For the case of interest here of fluid motion in flat space in 1+1 dimensions its tensor and scalar parts may be expanded as
\bse
\label{omitted}
\bal
\pi_{(2)}^{\mu\nu} &= \tilde c_1 \tilde{\mathcal{O}}_1 ^{\mu \nu}  + \tilde c_2 \tilde{\mathcal{O}}_2 ^{\mu \nu} + \tilde c_7 \tilde{\mathcal{O}}_7^{\mu\nu} \,, \\
\Pi_{(2)} &= \tilde b_2 \tilde{\mathcal{S}}_2 + \tilde b_3 
\tilde{\mathcal{S}}_3 + \tilde b_4 \tilde{\mathcal{S}}_4 \,.
\end{align}
\ese
In order to make contact with \cite{Attems:2017ezz} we chose the basis of operators  to be
\bse
\label{basis1}
\bal
\tilde{\mathcal{O}}_1^{\mu \nu} &= 2 \nabla^{<\mu} \nabla^{\nu>} 
\mathcal{E}  \sac \tilde{\mathcal{O}}_2^{\mu \nu} =  2  \nabla^{<\mu} \mathcal{E}  \nabla^{\nu>} 
\mathcal{E} \,, \\
\tilde{\mathcal{O}}_7^{\mu\nu}&= \frac{\nabla \cdot u}{3} \sigma^{\mu\nu}
\sac \tilde{\mathcal{S}}_2 = 2 \nabla_\mu \mathcal{E} 
\nabla^\mu \mathcal{E}
\\
\tilde{\mathcal{S}}_3 &=2  \nabla^2 \mathcal{E} \sac
\tilde{\mathcal{S}}_4 =  \left(\nabla \cdot u\right)^2 \,.
\end{align}
\ese  
Part of the notation above is chosen to make contact with 
\cite{Baier:2007ix,Romatschke:2009kr} below. The coefficients 
$\tilde c_1,  \tilde c_2,  \tilde b_2,  \tilde b_3$ are known because they are related to the coefficients $c_L, c_T, f_L, f_T$ determined in \cite{Attems:2017ezz}. We have not computed the coefficients $\tilde c_7, \tilde b_4$ but they are not needed in order to obtain a good hydrodynamic description. As in  \cite{Attems:2017ezz}, the reason is that the operators 
$\tilde{\mathcal{O}}_7^{\mu\nu}, \tilde{\mathcal{S}}_4$ are highly suppressed in the dynamical situation under consideration because they are quadratic in the fluid velocity, which is very small because the blob is quasi-static. This result is  illustrated in \fig{fig:pressures_midrapidity}, where we compare the exact pressures $P_L, P_T$ read off from gravity with the second-order hydrodynamic pressures $P_L^{\mt{hyd}}, P_T^{\mt{hyd}}$. To obtain the latter we read off the energy density and the fluid velocity from gravity and we apply the constitutive relations \eqq{stresstensor} with 
$\Pi_{(2)}^{\mu\nu}$ given by \eqq{omitted} omitting the contributions of 
$\tilde{\mathcal{O}}_7^{\mu\nu}$ and  $\tilde{\mathcal{S}}_4$. In \fig{fig:pressures_midrapidity}(top row) we see that the result agrees well with the exact pressures at mid rapidity after a hydrodynamization time 
$\thy \simeq 19.01/\Lambda \simeq 4.64 /T_\mt{hyd}$, 
with $T_\mt{hyd}$ the  temperature at mid rapidity at $t=\thy$.  After this time 90\% of the fluid energy is moving with velocity $v_z<0.1$, hence the term quasi-static. 
In fact, hydrodynamics  describes well not just the time dependence of the pressures at mid rapidity but also the spatial profile of the blob away from mid rapidity at sufficiently late times, as illustrated by \fig{fig:pressures_midrapidity}(bottom row). In \fig{fig:pressures_midrapidity} we have also plotted the ideal (equilibrium) pressure, as well as the hydrodynamic pressures obtained by including only the first-order viscous corrections. The fact that both agree with one another at late times but fail to describe the exact pressures shows that 
the first-order terms are suppressed 
and that  the second-order terms are as large as the ideal terms.

The purely spatial formulation of hydrodynamics is an acausal theory for which the initial-value problem is not well posed. For second-order hydrodynamics, a cure that is vastly used in hydrodynamic codes consists of  using the first-order equations of motion to exchange the terms with second-order purely spatial derivatives in the local rest frame for terms with one time and one spatial derivative (see \cite{Romatschke:2017ejr} for a review). This results in what we will call a M\"uller-Israel-Stewart-type (MIS) formulation. 
We emphasize that, strictly speaking, what is known as the MIS formulation is the phenomenological approach introduced in \cite{Muller:1967zza,Israel:1979wp,Israel:1976tn}, which is not second-order accurate. Building on it, different second-order accurate formulations have been constructed \cite{Baier:2007ix,Romatschke:2009kr,Denicol:2012cn}, to which we will collectively refer as MIS-type formulations. The key point is that, while they differ from MIS and they may also differ from one another in certain details, all these formulations share the common property that a second-order spatial derivative is replaced by one time and one spatial derivative as a first step to  to make the initial-value problem well posed. Since these two sets of second-order terms differ by higher-order terms, the purely spatial formulation and the MIS-type formulations are equivalent if all gradients are small \cite{inother}.
Since second-order gradients are large in our situation, one may expect that the two formulations will differ, 
as we will now verify.  
We follow \cite{Romatschke:2009kr}, which is completely general for a non-conformal neutral fluid (see  \cite{Baier:2007ix} for the conformal case).

In 3+1 dimensions the tensor and the scalar parts of $\Pi_{(2)}^{\mu\nu}$ can be expanded in a basis of eight tensor operators $\mathcal{O}_i^{\mu\nu}$  and seven scalar operators $\mathcal{S}_j$, respectively \cite{Romatschke:2009kr}. For the case of fluid motion in flat space in 1+1 dimensions only the following operators of the basis chosen in \cite{Romatschke:2009kr} do not vanish identically \cite{o1}:
\bse
\label{basis2}
\bal
\mathcal{O}_1^{\mu \nu} &= -2 c_s^2 \left(\nabla^{< \mu} \nabla^{\nu > } \log s - c_s^2 \nabla^{< \mu}  \log s \nabla^{\nu > }  \log s \right)\,, \,\,\,\,\,\,
\\
\mathcal{O}_3^{\mu\nu} &= \sigma^{<\mu}_\lambda \sigma^{\lambda \nu>}
\sac \mathcal{O}_7^{\mu\nu}=\tilde{\mathcal{O}}_7^{\mu\nu} \,,
\\
\mathcal{O}_8^{\mu\nu} &=  \nabla^{< \mu}  \log s \nabla^{\nu > }  \log s 
\sac \mathcal{S}_1 = \sigma_{\mu\nu} \sigma^{\mu \nu} \,,
\\
\mathcal{S}_3&= c_s^2 \nabla_\mu \nabla^\mu \log s + \frac{c_s^4}{2} \nabla_\mu \log s \nabla^\mu \log s + \frac{1}{6} \left(\nabla \cdot u\right)^2\,,
\\
{\mathcal{S}}_4 &=\tilde{\mathcal{S}}_4 \sac
\mathcal{S}_6= \nabla_\mu \log s \nabla^\mu \log s \,.
\end{align}
\ese
In these expressions $s$ is the entropy density and $\nabla^\mu \log s = s^{-1} \, \nabla^\mu s$.  Note that the ${\mathcal{O}}_7^{\mu\nu}$ and $\mathcal{S}_4$ operators are the same in both basis.  Moreover, in 1+1 dimensions we have 
\be
\label{madeuse}
\mathcal{O}_3^{\mu\nu}=2 \mathcal{O}_7^{\mu\nu} \sac 
\mathcal{S}_1=\frac{8}{3} \mathcal{S}_4 \,,
\ee
showing that the number of independent operators is the same as in \eqq{omitted}. The first-order equations of motion imply the following identities 
\bse
\label{replace}
\bal
&D \sigma^{\mu\nu} + \mathcal{O}_7^{\mu\nu} = \mathcal{O}_1^{\mu\nu} - \frac{1}{2} \mathcal{O}_3^{\mu\nu} - 2 \frac{d c_s^2}{d \log s} \mathcal{O}_8^{\mu\nu} \,,
\\
&D \left( \nabla \cdot u \right) =  -\frac{1}{4} \mathcal{S}_1 - \mathcal{S}_3 -\frac{1}{6} \mathcal{S}_4 +\left(\frac{3}{2} c_s^4 -  \frac{d c_s^2}{d \log s}\right) \mathcal{S}_6 \,,
\end{align}
\ese
where $D=u^\mu \del_\mu$ is the time derivative in the local rest frame and the equal signs here mean equality up to third- and higher-order terms. These identities may be used to replace $\mathcal{O}_1^{\mu\nu}$ and $\mathcal{S}_3$ in the expansions of $\pi_{(2)}^{\mu\nu}$ and $\Pi_{(2)}$ in favor of the left-hand sides of \eqq{replace} \cite{Romatschke:2009kr}, thus replacing terms with two spatial derivatives in the local rest frame for terms with one time and one spatial derivative. Upon these replacement the expansions read 
\bse
\label{omitted2}
\bal
\pi_{(2)}^{\mu\nu} &= \eta \tau_\pi \left(D \sigma^{\mu\nu} + 
\mathcal{O}_7^{\mu\nu}\right) + \left( 2 \lambda_1  + \eta \tau^*_\pi \right) \mathcal{O}_7^{\mu\nu}+ \lambda_4 \mathcal{O}_8^{\mu\nu} \,,
 \\
\Pi_{(2)} &=\zeta \tau_\Pi  D \left( \nabla \cdot u \right)  + 
\left( \frac {8}{3} \xi_1 +\xi_2 \right) \mathcal{S}_4+  \xi_4 \mathcal{S}_6 \,,
\end{align}
\ese
where we have made use of \eqq{madeuse} and we have labelled the second-order coefficients as in  \cite{Romatschke:2009kr}.
The coefficients in this expansion can be related to those in \eqq{omitted} by changing from one basis of operators to the other. Upon this change the fact that in our dynamical situation $\tilde c_7 \tilde{\mathcal{O}}_7^{\mu\nu} \simeq \tilde b_4 \tilde{\mathcal{S}}_4 \simeq 0$ translates into 
\bse
\label{reprep}
\bal
\left( 2 \lambda_1  +  \eta \tau_\pi^* \right) \mathcal{O}_7^{\mu\nu} &\simeq 
2\eta \tau_\pi \mathcal{O}_7^{\mu\nu} \,, \\
\left( \frac {8}{3} \xi_1 +\xi_2 \right) \mathcal{S}_4 & \simeq 
\zeta \tau_\Pi \mathcal{S}_4 \,.
\end{align}
\ese
Using \eqq{reprep} in \eqq{omitted2} we finally arrive at the MIS-type constitutive relations 
\bse
\label{omitted3}
\bal
\pi_{(2)}^{\mu\nu} &= \eta \tau_\pi  \left(D \sigma^{\mu\nu} + 
\mathcal{O}_7^{\mu\nu}\right) + 2 \eta \tau_\pi   \mathcal{O}_7^{\mu\nu}+ \lambda_4 \mathcal{O}_8^{\mu\nu} \,,
 \\
\Pi_{(2)} &=\zeta \tau_\Pi D \left( \nabla \cdot u \right)  + 
\zeta \tau_\Pi \mathcal{S}_4+  \xi_4 \mathcal{S}_6 \,.
\end{align}
\ese
As shown in \fig{fig:pressures_midrapidity}, the second-order hydrodynamic pressures  determined from these constitutive relations, $P_L^{\mt{hydMIS}}$ and  $P_T^{\mt{hydMIS}}$, fail to describe the exact pressures.

\noindent
{{\bf 5. Discussion.}}
We have seen that holographic collisions in the model \eqq{eq:action} result in the formation of a long-lived, quasi-static blob regardless of the order of the transition. In other words, out-of-equilibrium physics smooths out the details of the transition. We emphasize that this is a statement about the one-point function of the stress tensor. It would be interesting to investigate whether two-point functions  exhibit  qualitative differences between the case of a second-order phase transition, in which fluctuations are expected to be enhanced \cite{Stephanov:1998dy,Stephanov:1999zu},  and the other two cases.

We emphasize that our model differs from QCD in important respects. The conjectured critical point in QCD lies at non-zero baryon chemical potential, meaning that the physics near the critical point involves a dynamical baryon charge density. Moreover, the critical point in QCD would be characterised by a conserved order parameter. Both of these features are absent in our model, in which there is no dynamical baryon charge and no conserved order parameter. Despite these differences, however, it is plausible that our results may hold qualitative lessons for the dynamics near a putative critical point in QCD. The reason is that the formation of the blob seems to be due to a property that is shared by any critical point, namely the vanishing of the speed of sound. Indeed, the freezing of the blob dynamics, which is ultimately responsible for the failure of the MIS-type hydrodynamics,  seems to be caused by the fact that the speed of sound in absolute value is small at the energy densities of the blob, $|c_s^2| \simeq 10^{-2}$, as can be seen from Figs.~\ref{fig:energyVStemperature0} and \ref{fig:EnergyScanCriticalPoint}. If  $\phi_M \gg \phi_M^*$ the crossover is no longer a rapid one and the minimum of $c_s^2$ is no longer small. Consistently, in this case no blob is formed \cite{Attems:2016tby,Attems:2017zam}. It would be interesting to investigate if the blob formation persists in theories with $\phi_M \ll \phi_M^*$ in which the first-order phase transition is stronger and the minimum of $c_s^2$ is negative but again not small in absolute value. It would also be interesting to allow for non-trivial dynamics in the transverse plane, since  regions with negative $c_s^2$ should suffer from a spinodal instability with the consequent formation of inhomogeneities \cite{Attems:2017ezz}.  

At a time such that $\thy \Thy \sim \mathcal{O}(1)$ the time evolution of the blob at mid rapidity becomes well described by the purely spatial  formulation of  hydrodynamics. As time progresses the spatial profile of the blob  becomes well described too. We emphasize that this  early hydrodynamization is only achieved when second-order terms are included. Presumably the agreement with first-order hydrodynamics would eventually occur at much later times, in accord with \cite{Critelli:2018osu}.

Purely spatial hydrodynamics is known to be acausal. This was not an issue for us since we did not evolve in time the hydrodynamic equations but simply verified the constitutive relations, but it is an issue in situations in which hydrodynamics is the only available description.  In the MIS-type formulation, acausality is remedied by replacing terms with second-order spatial derivatives in the local rest frame by terms with one time and one space derivative. In the limit of small gradients this produces an equivalent formulation at long wavelengths. However, in our situation the result is not equivalent and in fact it fails to describe the correct evolution of the stress tensor even at late times 
such that $tT\gg 1$. 
This is not surprising since the quasi-static, inhomogeneous blob has small time gradients but large spatial gradients. 
Although deviations from MIS-type formulations for large gradients have been reported previously \cite{Florkowski:2013lya}, for fluids with small viscosity those were limited to early-time evolution. 
Note that, strictly speaking, the MIS formulation would require a second step in which the tensor $\Pi^{\mu\nu} $ is promoted to a dynamical variable. We have not considered this second step since the first one already produces an inequivalent description.  

The success of hydrodynamics in the presence of large spatial gradients has been noted before  \cite{Chesler:2010bi,Casalderrey-Solana:2013aba,Casalderrey-Solana:2013sxa,Chesler:2015wra,Chesler:2015bba,Chesler:2015lsa,Chesler:2016ceu, Heller:2011ju,Kurkela:2015qoa}.
In all those cases first-order viscous corrections were as large as the ideal terms, and the combination of the two produced a good  description of the flow. In contrast, in our dynamical simulation first-order gradients are suppressed, and the leading gradients are the second-order terms,  which become comparable to the ideal terms. The agreement with the exact pressures  implies that the sum of all higher-order gradient contributions is small. However, this crucially depends on having organized the hydrodynamic expansion in terms of purely spatial gradients. 
The failure of the MIS-type formulation to describe the microscopic dynamics shows that for other choices of second-order gradients those additional higher-order terms cannot be neglected. 
The inequivalence of the different second-order choices has also been observed in the limit of large number of dimensions, where only the purely spatial formulation becomes exact \cite{Andrade:2018zeb}  without introducing field redefinitions \cite{Herzog:2016hob,Rozali:2017bll}. 

It would be extremely interesting to develop a new causal formulation of hydrodynamics that includes  the necessary purely spatial gradients. This could have a tremendous impact on the current world efforts for locating the QCD critical point. Our results suggest that hydrodynamic codes, which are based on an MIS-type formulation, may need to be modified in order to describe the early-time evolution of high-density heavy ion collisions. Note also that, near a critical point, the slow dynamics associated to the long-lived fluctuations of the order parameter lead to modifications of hydrodynamics \cite{Stephanov:2017ghc} complementary to those that we have described. In our model, the order parameter is a combination of the energy density,  
$\mathcal{E}$, and the expectation value of the scalar operator, $\Lambda \langle \mathcal{O} \rangle$. Both of these quantities jump discontinuously across the first-order phase transition and their susceptibilities diverge at the second-order phase transition. Correspondingly, on the gravity side fluctuations of the metric mix with those of the scalar field and the  
 mode whose correlation length diverges at the critical point is a linear combination of these fluctuations. However, the effects of this single new soft  mode are $1/\nc^2$-suppressed with respect to those in the $O(\nc^2)$ hydrodynamic stress tensor \cite{Natsuume:2010bs}, and are therefore postponed to parametrically later times. Understanding the interplay of these two type of modifications in QCD may be essential to correctly interpret high-baryon density heavy ion data.

\noindent
{{\bf Acknowledgements.}} 
We thank M.~Heller, C.~Herzog, P.~Romatschke, M.~Spali\'nsky, M.~Stephanov and 
M.~Strickland for discussions. We thank the MareNostrum supercomputer at the BSC for computational resources (project no.~UB65). Computations were also performed on the Baltasar cluster at IST.  JCS is a Royal Society University Research Fellow (on leave). MA acknowledges support through H2020-MSCA-IF-2014 FastTh 658574, PA2014-58293-C2-1-P, and MDM-2016-0692. MZ acknowledges support through the FCT (Portugal) IF programme, IF/00729/2015. We are also supported by grants FPA2016-76005-C2-1-P, FPA2016-76005-C2-2-P, 2014-SGR-104, 2014-SGR-1474, SGR-2017-754 and MDM-2014-0369.




\begin{thebibliography}{99}
\bibitem{inother}
In principle, both a purely spatial and an MIS-type formulation can be consistently carried out to arbitrary order in the derivative expansion. The difference between them can be viewed as a reorganization of the different terms in the effective theory that hydrodynamics is. In all cases the hydrodynamic equations of motion are obtained by applying to the corresponding expansion  the conservation of the gauge theory stress tensor, $\partial_\mu T^{\mu\nu}$, which one may regard as the Ward identity associated to translation invariance in the gauge theory. The first-order equations are those obtained by truncating these expansions to first order.
\end{thebibliography}

\end{document}